\documentclass[%
 reprint,
nofootinbib,
 amsmath,amssymb,
 aps,
]{revtex4-2}

\usepackage{graphicx}
\usepackage{dcolumn}
\usepackage{bm}

\def\prd{Phys. Rev. D}

\def\mnras{Monthly Notices of the Royal Astronomical Society}
\def\apj{The Astrophysical Journal}

\def\prd{Phys. Rev. D}

\def\mnras{Mon. Not. R. Astron Soc.}
\def\apj{The Astrophysical Journal}

\providecommand{\dif}{\mathrm{d}} 

\begin{document}

\preprint{APS/123-QED}

\title{Electric Penrose process: high-energy acceleration of ionized particles by \\ non-rotating weakly charged black hole 
}

\author{Arman Tursunov}
 \email{arman.tursunov@physics.slu.cz}
\author{Bakhtinur Juraev}%
\author{Zden{\v e}k Stuchl{\' i}k}
\author{Martin Kolo{\v s}}
\affiliation{%
Research Centre for Theoretical Physics and Astrophysics, Institute of Physics, Silesian University in Opava, Bezru\v{c}ovo n\'{a}m.13, CZ-74601 Opava, Czech Republic
}%


\begin{abstract}
In many astrophysical scenarios the charge of the black hole is often neglected due to unrealistically large values of the charge required for the Reissner-Nordstr\"om spacetime metric. However, black holes may possess small electric charge due to various selective accretion mechanisms. In this paper we investigate the effect of a small hypothetical electric charge of a Schwarzschild black hole on 
the ionization of a freely falling neutral particle and subsequent escape of the ionized particle from the black hole. We show that the energy of ionized particle can grow ultra-high and discuss distinguishing signatures of particle acceleration by weakly charged black holes. We also discuss a possible application of the proposed mechanism as an alternative cosmic ray acceleration scenario. In particular we show that the Galactic centre supermassive black hole is capable to act as a PeVatron of protons. The presented mechanism can serve as a simple toy model of a non-rotating compact object acting as a particle accelerator with a potential astrophysical implementations related to the cosmic ray  physics and beyond. 
\end{abstract}

\maketitle


\section{Weakly charged black hole}\label{intro}

Recently, it has been pointed out that the ionization or decay of neutral particles in the vicinity of rotating Kerr black hole immersed into external magnetic field can lead to the acceleration of ionized particles to ultra-high energies, with the Lorentz $\gamma$-factors of particles that may exceeding $10^{12}$ near supermassive black holes in realistically plausible conditions \citep{2020ApJ...895...14T,2019Univ....5..125T}. The formalism of the mentioned acceleration mechanism is based on the magnetic Penrose process \citep{Wag-Dhu-Dad:1985:APJ:,Par-etal:1986:APJ:} in its novel, ultra-efficient regime, in which the energy of ionized particle drives away the rotational energy of the black hole through electromagnetic interaction. It has been claimed that the mechanism might be responsible for the production of the highest-energy cosmic ray particles  \citep{2020ApJ...895...14T} with energy exceeding $10^{20}$eV when applied to realistic supermassive black hole candidates. The process, however, requires the rotation of the black hole and the presence of external magnetic field. 

In this paper we investigate whether the acceleration of ionized particles can be achieved in a more simplified settings, namely, in the vicinity of a non-rotating Schwarzschild black hole with a radial test electric field. By a \textit{test} electric field we denote the field, whose energy-momentum tensor can be neglected in the description of the gravitational field of the black hole. This implies that the electric field influences on the dynamics of charged particles only, being negligible for the geodesics of neutral particles. Such a simplified setup is motivated by the following reasons.

First of all, the \textit{no-hair theorem} of black hole physics states that the spacetime around black holes can be fully described by at most three metric parameters -- black hole mass, spin and electric charge \cite{Misner73}. The later is usually neglected in astrophysical scenarios justified, on the one hand, 
by unrealistically large values of the charge required for its visible effect on the spacetime metric and on the other hand by quick discharge of any charge 
excess by an accretion of a plasma surrounding black hole.  Indeed, one can compare the gravitational radius of a black hole with the characteristic length of the charge $Q_{\rm G}$ of the Reissner-Nordstr\"om black hole, which gives the maximum charge of the black hole per solar mass\footnote{Hereafter in this section  we use the esu-cgs system of units, in which the electrostatic unit of charge is given by 1 Fr $\equiv$ 1 esu = 1 cm$^{3/2}$ g$^{1/2}$ s$^{-1}$. In SI system of units, 1 C $= 3 \times 10^9$ Fr.}
\begin{eqnarray}
&&\sqrt{\frac{Q_{\rm G}^2 G}{c^4}} = \frac{2 G M}{c^2}, \\
  &\Rightarrow&  Q_{\rm G} = 2 G^{1/2} M \approx 10^{30} \frac{M}{M_{\odot}} ~\mbox{Fr}. \label{RNcharge}
\end{eqnarray}
%
%
%
This value of the charge is unattainable in any known astrophysically relevant scenario. 
Moreover, a neutralization of such a hypothetical charge $Q_{\rm G}$ would require an accretion of a net charge with the total mass of accreted charged particles of $M_Q = m_{p,e} Q_{\rm G} /e$,  where the indices 
denote proton and electron, respectively, and $e$ is an elementary charge. 
Luminosity of black hole surrounded by plasma or accretion disk can be derived from infalling matter as $L = \epsilon \dot{M} c^2$, where $\dot{M}$ accretion rate and $\epsilon$ is the fraction of the rest mass energy radiated away. On the other hand, from the balance of gravitational force and radiation pressure in the vicinity of a black hole one can derive the Eddington luminosity for fully ionized hydrogen plasma surrounding a black hole in the form
\begin{equation}
    L_{\rm Edd} = \frac{4\pi G M m_p c}{\sigma_{\rm T}} \approx 1.26 \times 10^{38} \left(\frac{M}{M_{\odot}}\right) {\rm erg/s}.
\end{equation}
Defining  charged matter accretion rate as the fraction of total accretion rate, $\dot{M}_Q = \delta \cdot \dot{M}$, one can derive the neutralization timescale of the maximally charged black hole in the following form
\begin{equation}
t_{Q, {\rm acc.}} = \frac{4}{3} \frac{e^3 \, \epsilon \, m_{p,e} }{ G^{1/2} \,c^3 \, \delta \,  m_p m_e^2} \approx 2.5 \times 10^{-2} \left( \frac{m_{p,e}}{m_p} \right) \left(\frac{\epsilon}{\delta}\right) {\rm s},
\end{equation}
which is estimated for a positive black hole charge. In case of a negative charge of the black hole, the timescale is $\approx 1835$ times faster. Both ${\epsilon}$ and $\delta$ have values in the range $(0\, , \, 1)$ and in many cases are of similar orders of magnitude. This implies that in all  astrophysically relevant settings any net charge of a black hole would be neutralized relatively quickly unless there is a mechanism preventing the black hole from neutralization. Some of such mechanisms are briefly mentioned below. Thus, the Reissner-Nordstr\"om spacetime metric is interesting, but astrophysically not viable. An exception can be represented by the Reissner-Nordstr\"om spacetime having a zero electric and non-zero magnetic charge due to special character of interaction with electrically charged matter \cite{1983BAICz..34..129S}. 

There exist several astrophysical scenarios based on a selective accretion, in which an astrophysical black hole may possess a small electric charge \cite{1975PhRvD..12.2959R,1978ApJ...220..743B,2006ApJ...645.1188W,Wald:1974:PHYSR4:,2019Obs...139..231Z,2018MNRAS.480.4408Z,2018PhRvD..98l3002L,2020ApJ...897...99T}. Since protons are about 1836 times more massive than electrons, the balance between the gravitational and Coulombic forces for the particles close to the surface of the compact object is obtained when the black hole acquires a positive net electric charge of the order of $Q \sim 3 \times 10^{11}$Fr per solar mass \citep{2019Obs...139..231Z,1978ApJ...220..743B}. Moreover, matter surrounding black hole can be ionized and charged by the irradiating photons taking away some electrons \citep{2006ApJ...645.1188W}. In that case, following the above described argument, the charge of the black hole is likely positive, being of the order of $Q \sim 10^{11}$Fr per solar mass.  Another famous mechanism of charging of black holes is the Wald's mechanism \cite{Wald:1974:PHYSR4:}, in which the charge is naturally induced by the twisting of magnetic field lines due to the frame-dragging effect of the rotation of black hole. As a result, both the black hole and surrounding magnetosphere should acquire equal and opposite charge of the order of $Q \sim 10^{18}$Fr per solar mass \citep[see, e.g.][]{1975PhRvD..12.2959R,Wald:1974:PHYSR4:,2016PhRvD..93h4012T}. In all cases, the charge of the black hole is much weaker than its maximal theoretical limit (\ref{RNcharge}) by many orders of magnitude. Therefore,  astrophysical black holes can be considered as weakly charged, i.e. gravitational effect of the charge on the spacetime metric can be rightly neglected. Thus, depending on whether the black hole is spinning or not, the realistic value of the black hole's charge may vary between the values 
\begin{equation}
    10^{11} \frac{M}{M_{\odot}} \, {\rm Fr} \lesssim Q_{\rm BH} \lesssim 10^{18} \frac{M}{M_{\odot}} \, {\rm Fr}. \label{BHchargelimits} 
\end{equation}
 For more details and estimates, the reader may refer to the works \cite{2018MNRAS.480.4408Z,2019Obs...139..231Z} and references therein, where various black hole charging scenarios are compared and the results are applied to the Galactic centre black hole.

The presence of induced electric field and corresponding electric charge of a black hole plays an important role in the mechanisms of the rotational energy extraction from black holes, 
such as Blandford–Znajek mechanism \cite{Bla-Zna:1977:MNRAS:} and magnetic Penrose process \cite{2019Univ....5..125T}. Discharge of the induced charge by accreting charged matter drives away the black hole's rotational energy. The character of the electric field around magnetized Kerr black hole is not spherically symmetric, it is rather of quadrupole character. Therefore, in most of the realistic particle acceleration scenarios, such as the production of relativistic jets, it is assumed that the black hole is spinning.

In this paper, we raise a question whether the charged particles can be accelerated to large $\gamma$-factors $>>1$ by non-spinning Schwarzschild black holes or in the case when the rotation of the black hole is very slow for the operation of the rotational energy extraction mechanisms? For that we assume that the black hole is non-rotating with a small electric charge of above described limits, whose gravitational contribution is negligible. We are aimed to calculate the energy of the charged particle after the ionization of a neutral particle in the black hole's vicinity. 



Hereafter we use the signature $(-,+,+,+)$, and the system of geometric units, in which $G = 1 = c$, unless the constants are written explicitly and given in esu-cgs system of units. 

\section{Dynamics of charged particle }

\subsection{Background setup \& equations of motion}

We start from the Schwarzschild spacetime metric 
\begin{equation} 
ds^2 = -f(r) dt^{2} + f^{-1}(r)dr^2 + r^2(d\theta^2 +\sin^2{\theta} d \phi^2),
\label{metric}
\end{equation}
where $f(r)$ is the lapse function parametrized by the black hole mass $M$ as follows
\begin{equation}
   f(r) = 1 - \frac{2M}{r}.
\end{equation}
Let us assume the presence of the radial electric field with corresponding small electric charge $Q$ at the center of the coordinate. In this case, the only non-zero covariant component of the electromagnetic four-potential $A_{\mu} = (A_{t},0,0,0)$ has the following simple form
\begin{equation}
A_{t} = - \frac{Q}{r}.
\label{electric-potential}
\end{equation}
%
The anti-symmetric tensor of the electromagnetic field $F_{\alpha \beta} = A_{\beta, \alpha} - A_{\alpha, \beta}$ has the only one independent nonzero component
\begin{equation}
    F_{tr} = -F_{rt} = - \frac{Q}{r^2}. 
\end{equation}

\begin{figure*}
\centering
\includegraphics[width=0.32\hsize]{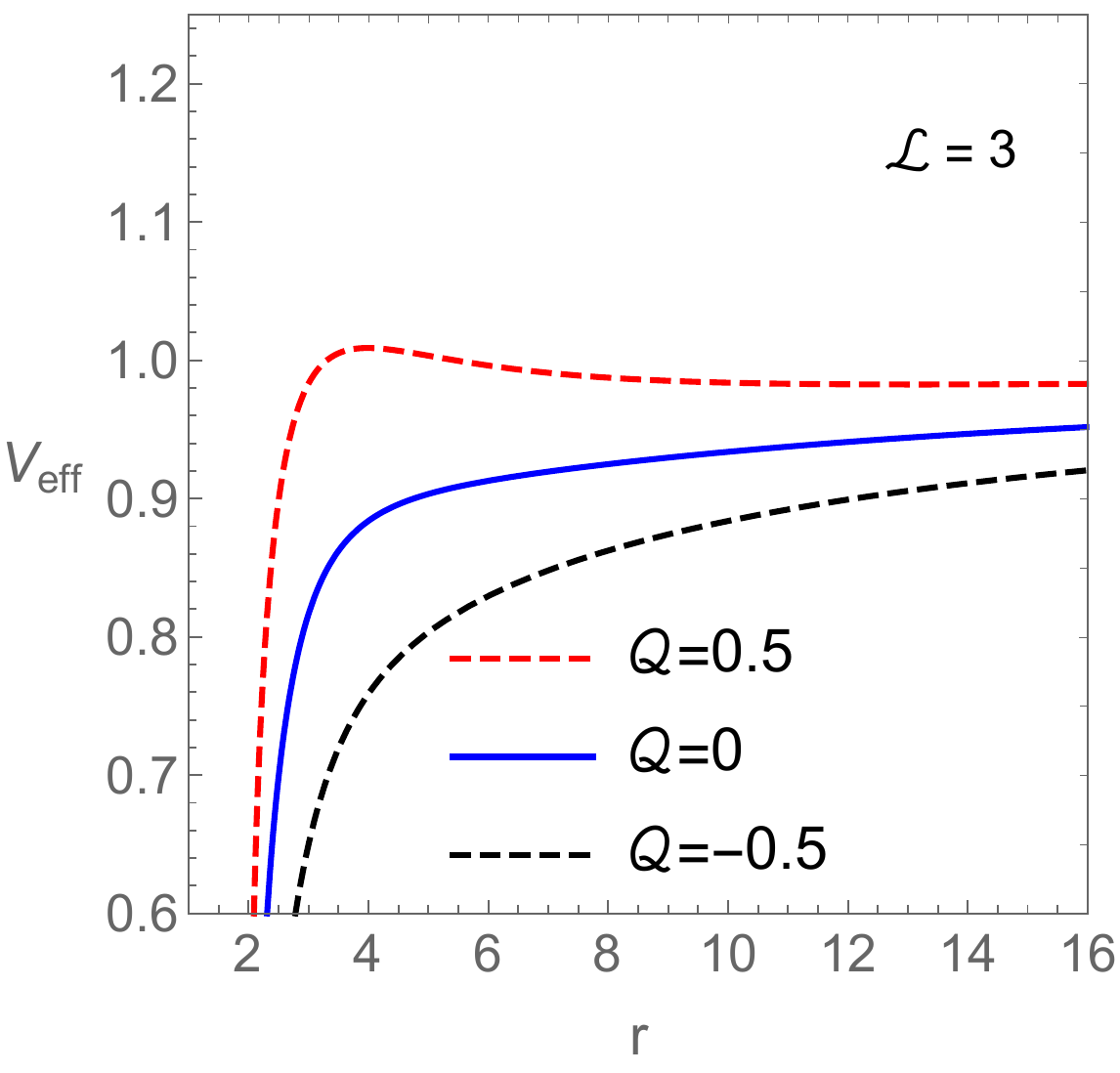}
\includegraphics[width=0.32\hsize]{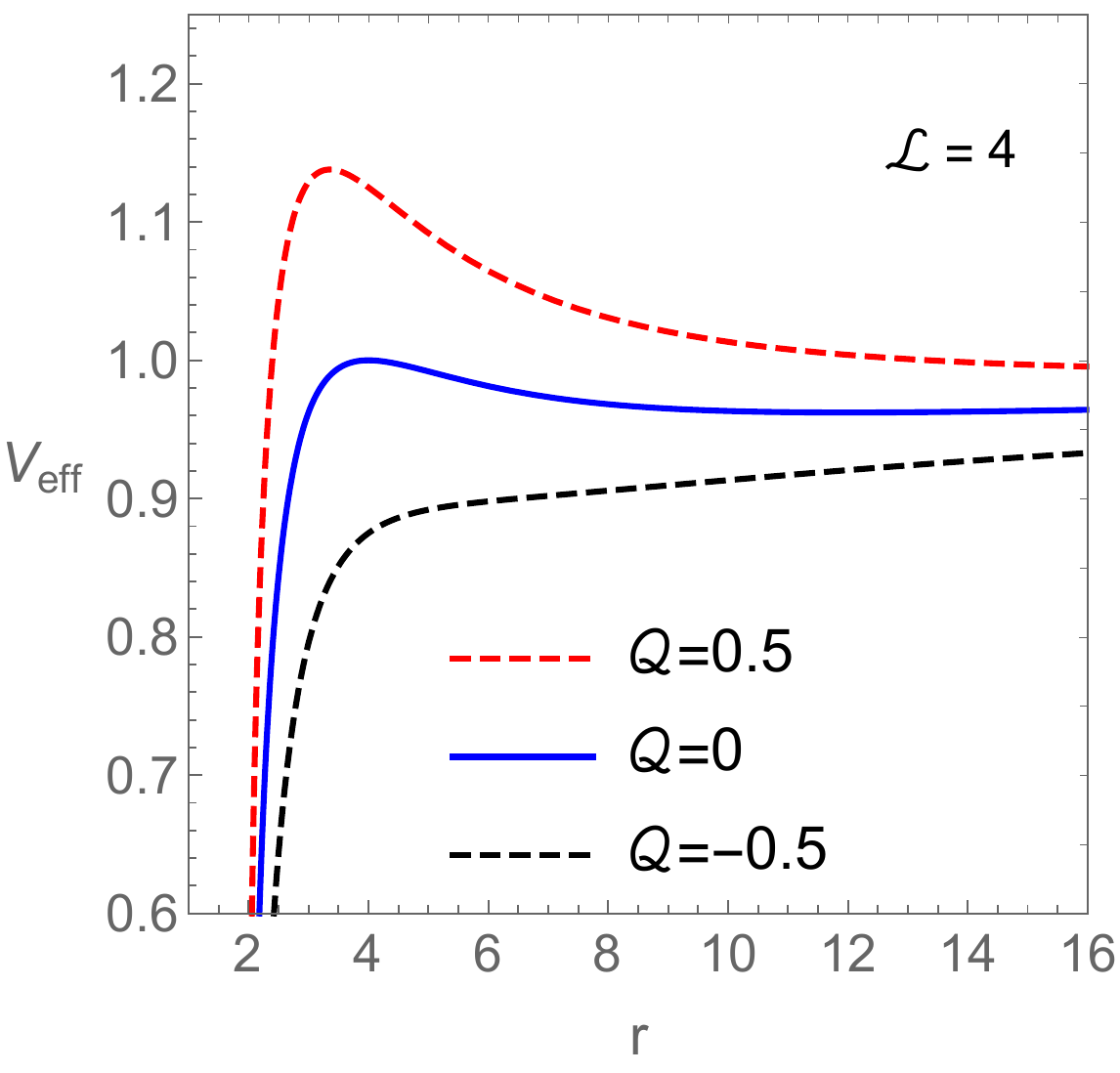}
\includegraphics[width=0.32\hsize]{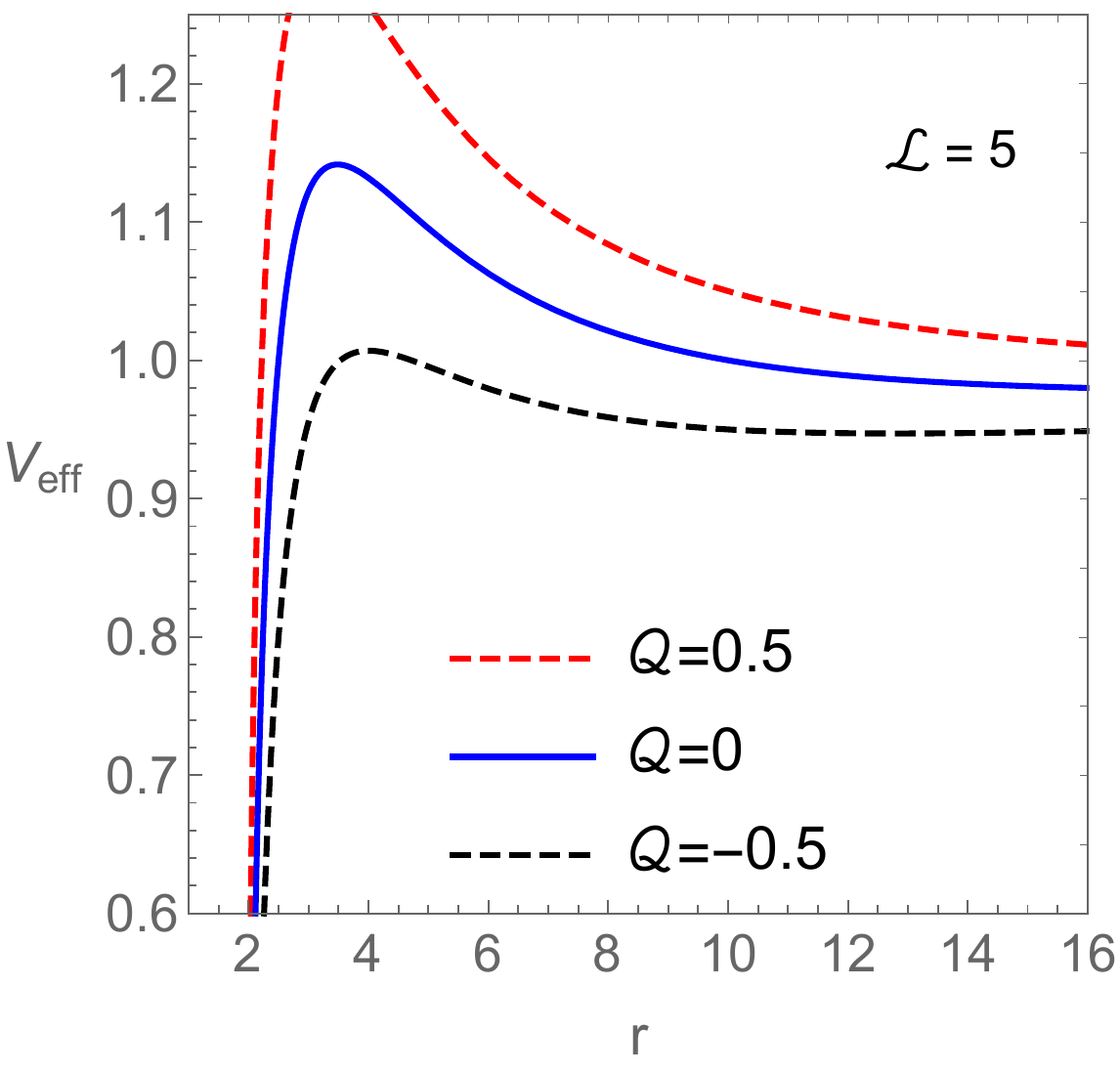}
\caption{Radial dependence of the effective potential $V_{eff}$ for a charged particle around weakly charged non-rotating black hole in the equatorial plane $\theta=\pi/2$ for different values of the parameters ${\cal L}$ and ${\cal Q}$.
\label{fig1}}
\end{figure*}

Let us now consider the motion of a charged particle of mass $m$ and charge $q$ in the combined background gravitational and electric fields. The motion of a charged particle is governed by the Lorentz equation in curved spacetime
%
%
\begin{equation} \label{Lorentzequ}
    \frac{\dif u^{\mu}}{\dif \tau} + \Gamma^{\mu}_{\alpha \beta} u^{\alpha} u^{\beta} - \frac{q}{m} F^{\mu}_{\,\,\,\nu} u^{\nu} = 0, 
\end{equation}
where $u^{\mu}$ is the four-velocity of the particle, $\tau$ is the proper time of the particle and $\Gamma^{\mu}_{\alpha \beta}$ -- Christoffel symbols.

Due to symmetries of the background Schwarzschild metric one can introduce two integrals of motion, corresponding to temporal and spatial components of the canonical four-momentum of the charged particle $P_\alpha = m u_\alpha + q A_\alpha$: 
\begin{eqnarray}
  \frac{P_{t}}{m} &=& - {\cal E} \equiv - \frac{E}{m}  = u_t - \frac{q Q}{m r}, \\
    \frac{P_{\phi}}{m} &=& {\cal L} \equiv \frac{L}{m} =  u_\phi,
\end{eqnarray}
where ${\cal E}$ and ${\cal L}$ denote specific energy and specific angular momentum of the charged particle. 
Since both gravitational and electric fields are spherically symmetric and there is no preferred plane of the motion, one can fix the motion of the charged particle to the equatorial plane ($\theta = \pi/2$), without loss of generality. Thus, three non-vanishing components of the equation of motion (\ref{Lorentzequ}) can be found in the form
\begin{eqnarray}
\frac{\dif u^t}{\dif \tau} &=& 
\frac{u^r \left[ Q r - 2 M (e r + Q)\right]}{r (r-2 M)^2} \label{eqmot}
\\
    \frac{\dif u^r}{\dif \tau} &=& 
     \frac{e Q}{r^2} + \frac{\mathcal{L}^2 (r-2 M)}{r^4}
     - \frac{M \left[e^2- (u^r)^2\right]}{r (r-2 M)},
\\
\frac{\dif u^\phi}{\dif \tau} &=& - \frac{2~\mathcal{L}~u^r}{r^3}, \label{eqmophi}
\\
 {\rm where} && e = {\cal E} - \frac{q Q}{m r}.
\end{eqnarray}
Equations (\ref{eqmot}) - (\ref{eqmophi}) are ordinary differential equations, which can be easily solved numerically. 

%


\subsection{Effective potential}

Using the normalization condition for a massive particle $u^\mu u_\mu = -1$, one can derive the effective potential for the charged particle moving around a weakly charged Schwarzschild black hole in the form
\begin{equation}
V_{\rm eff}(r) =   \frac{\mathcal{Q}}{r} + \sqrt{f(r) \left( 1 + \frac{\mathcal{L}^2}{r^2} \right)},  
\label{Veff:Electric charge}
\end{equation}
where $\mathcal{Q} = Q q/m $ is a parameter characterizing the electric interaction between the charges of the particle and black hole. Without loss of generality we set the mass of the black hole to be equal to unity, i.e. $M=1$. 


Since the right hand side of the effective potential \eqref{Veff:Electric charge} is always positive one can distiguish two qualitatively different situations depending on the sign of the parameter $\mathcal{Q}$. When $\mathcal{Q}>0$, the charges of the particle and black hole have the same sign, so the electric interaction is repulsive. In opposite case, when $\mathcal{Q}<0$, the charges of the particle and black hole have different signs, so the electric interaction is attractive. 
The term ${\cal L}^2$ under the root of Eq.\eqref{Veff:Electric charge} means that the clockwise and counter-clockwise directions of the motion are equivalent.




The radial profile of the effective potential is shown in Figure \ref{fig1}. One can see that the effect of the charge parameter $\cal Q$ is similar to those of the angular momentum $\cal L$, i.e. increasing (or decreasing) both parameters $\cal Q$ and $\cal L$ one can increase (or decrease) the value of the effective potential. It is interesting to note that taking into account the parameter $\cal Q$ can mimic the effect of angular momentum (compare, e.g. red curve in the middle plot with a very similar blue curve on the right plot of the Figure \ref{fig1}).

\begin{figure*}
\centering
\includegraphics[width=0.32\hsize]{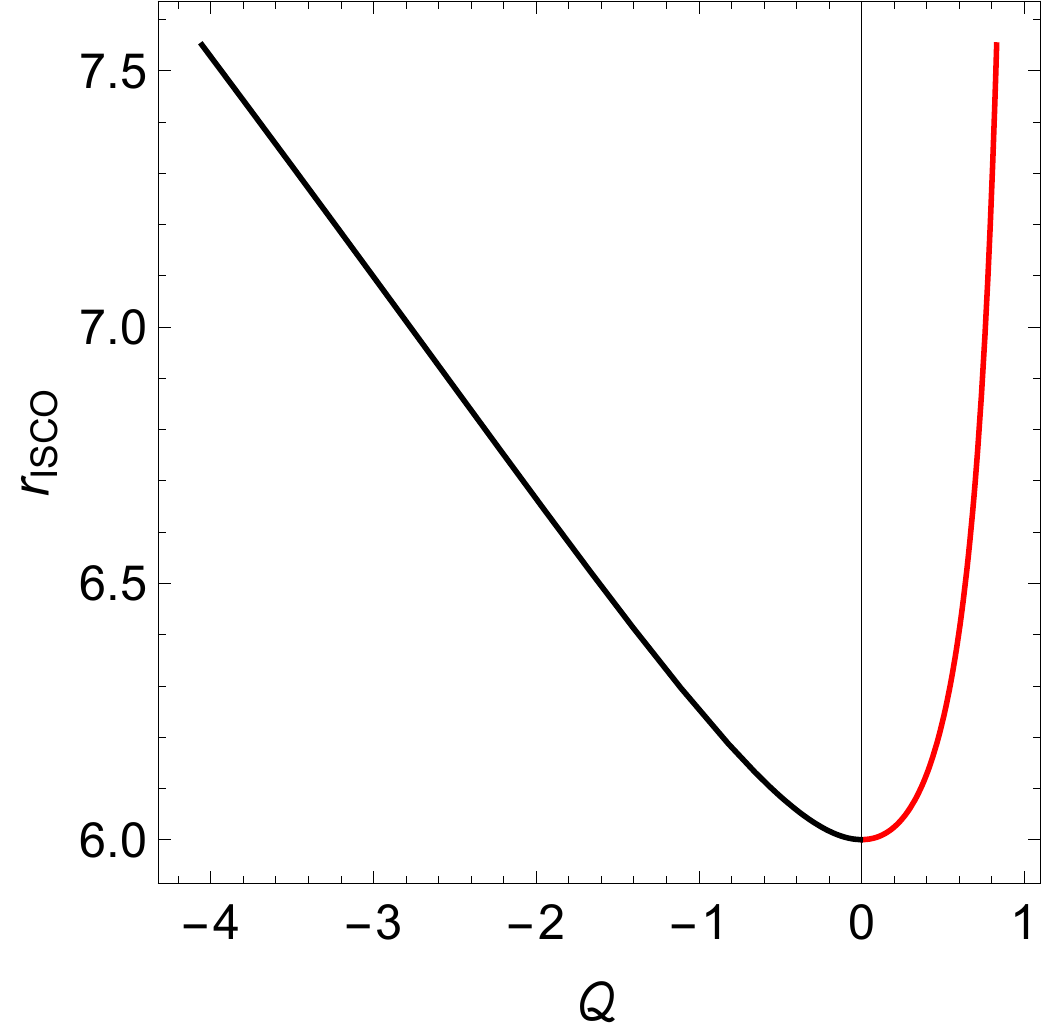}
\includegraphics[width=0.32\hsize]{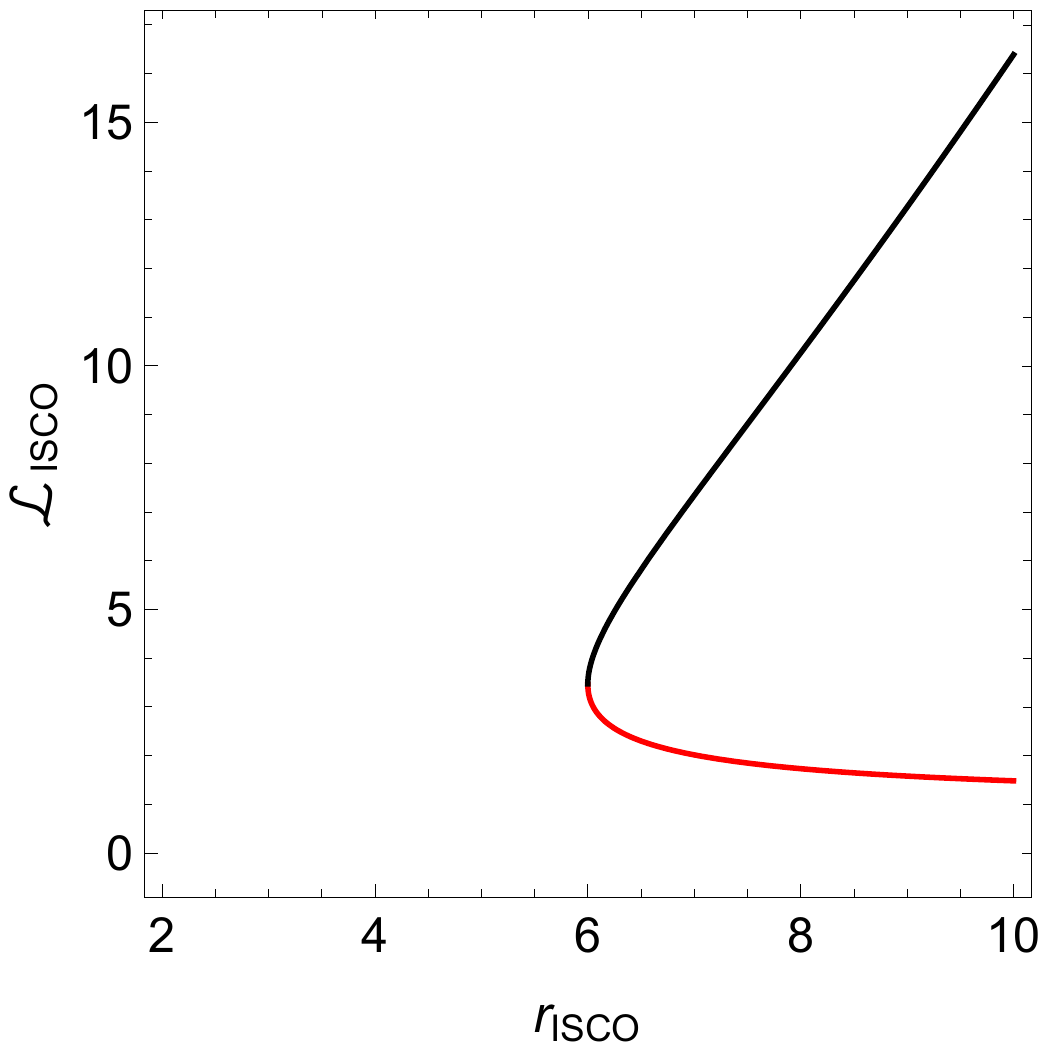}
\includegraphics[width=0.32\hsize]{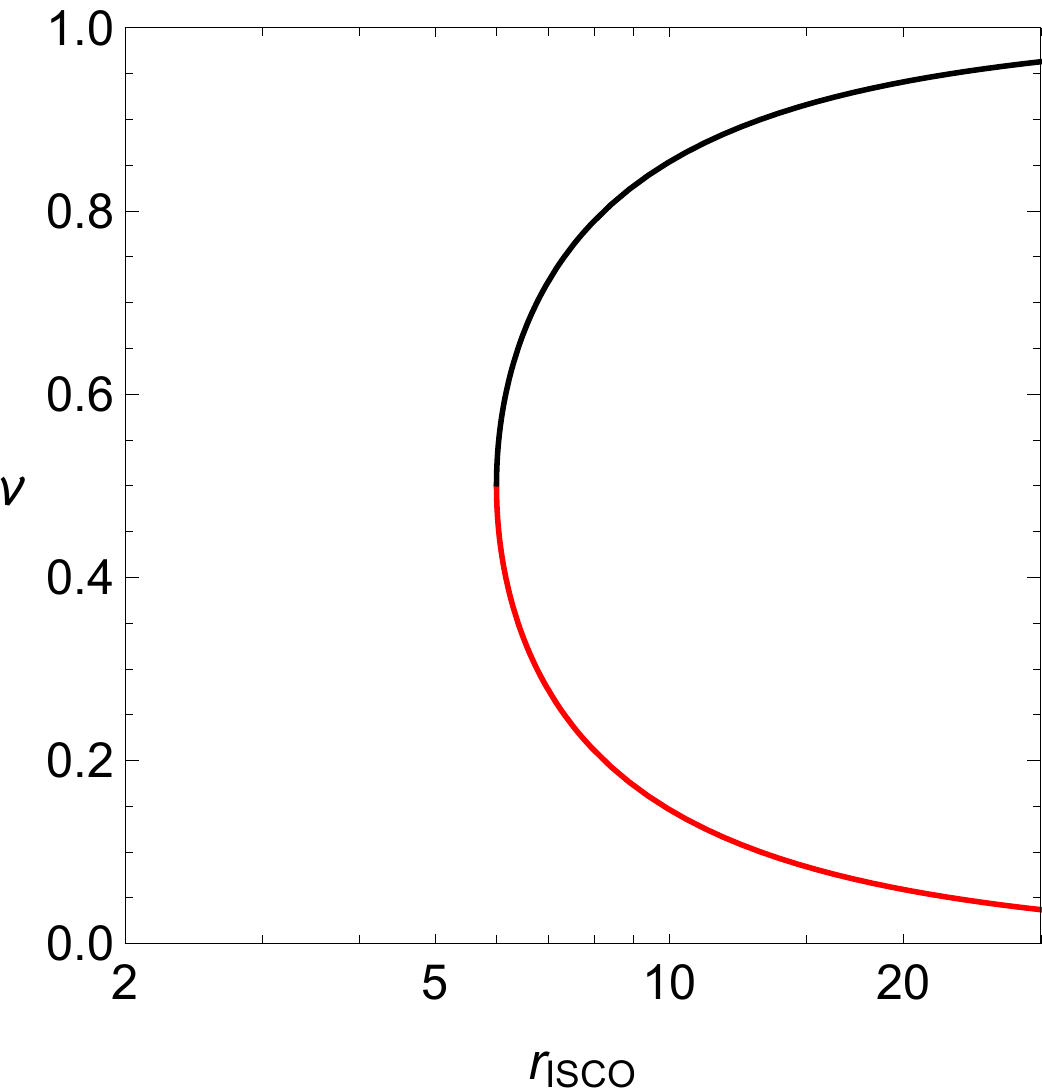}
\caption{Left: position of the ISCO of charged particle in the dependence on the charge parameter $\cal Q$. Middle: Angular momentum of charged particle at ISCO against ISCO position. Right: velocity of charged particle at ISCO. 
In all plots the red lines correspond to the positive charge parameter ${\cal Q}>0$, while black curves correspond to the negative charge parameter ${\cal Q}<0$. 
\label{fig2}}
\end{figure*}

The stationary points of the effective potential $V_{\rm eff} (r)$ is given by the equation
\begin{equation}
    \partial_{r} V_{\rm eff}(r)= 0. 
    \label{Veff:ISCO} 
\end{equation}
Note that in the case of the weakly charged Schwarzschild black hole all the local extrema of the effective potential $V_{\rm eff}$ are located in the equatorial plane $\theta = \pi/2$.
Equation \eqref{Veff:ISCO} leads to a polynomial equation of the fourth order in the radial coordinate
\begin{equation}
    r^2 ( J -1) + \mathcal{L}^2 (r-3) = 0,
    \label{der:veff,r}
\end{equation}

\begin{equation}
   \mbox{where} \quad  J= \frac{\mathcal{Q}}{r} \sqrt{\frac{(r-2) (\mathcal{L}^2 + r^2)}{r}}.
\end{equation}
The solution of equation \eqref{der:veff,r} has four roots of $\mathcal{L}$ and two of them being independent
\begin{equation}
\begin{split}
 \mathcal{L}_{\pm}^2 = \frac{r}{(r-3)^{2}}   \Big[ - \mathcal{Q}^2  - 3r + \frac{\mathcal{Q}^2 r}{2} + r^2 \\
 \pm \mathcal{Q} \sqrt{\mathcal{Q}^2 - 12r + 4 r^2} \left( 1 - \frac{r}{2} \right) \Big], 
 \end{split}
\end{equation}
%



\subsection{Angular velocity measured at infinity}

Noticing that in the equatorial plane the four velocity takes the form $u^{\alpha} = u^{t} (1,v,0,\Omega)$, where $v = dr/dt$, $\Omega = d\phi/dt$ and using the normalization condition $ u^{\alpha} u_{\alpha} =-k$, where $k=1$ for massive particle and $k=0$ for massless particle, we can obtain the following equation
\begin{equation}
    (u^{t})^2 (f^{-1}(r) v^{2} - f(r) + \Omega^2 r^2) = - k .
\end{equation}
Simplifying equation above, we can easily derive equation for angular velocity measured by a static observer at infinity $\Omega = d\phi/dt$
\begin{equation}
\Omega = \pm \frac{1}{u_{t} r} \sqrt{(u_{t})^2(f(r) - f^{-1}(r) v^2) - k f^2(r) } .
\label{AngularVelocity}
\end{equation}
%
The possible values of $\Omega$ are limited to
\begin{equation} \label{AngularVelocity:limits}
    \Omega_{-} \leq \Omega \leq \Omega_{+}, \quad \Omega_{\pm} = \pm \frac{\sqrt{f(r)}}{r}.
\end{equation}
corresponding to the photon motion.

\subsection{Innermost stable circular orbit}

The innermost stable circular orbit (ISCO) in the Schwarzschild spacetime is located at $r_{\rm ISCO} = 6M$. In the case when the electric charge is included, it will be shifted from $r_{\rm ISCO} = 6M$. Local extremum of the function $\mathcal{L}_{\pm}$ determine the innermost stable circular orbit (ISCO), its radius, angular momentum and energy. The ISCO can also be found from the condition of $\partial^2_{r} V_{\rm eff}(r, \mathcal{L} , \mathcal{Q}) = 0$, which gives
\begin{equation}
\begin{split}
   \mathcal{L}^2 r^2 (J (r-2)+2)+r^4 (J (r-2)-r+3) \\ +\mathcal{L}^4 ((r-3) r+3)=0. 
\end{split}
\end{equation}
Solving this equation with respect to $r$ gives us four solutions for the ISCO with only two of them being real and independent. 

One can also calculate the velocity $v$ of the charged particle at the ISCO, which is given by the formula
\begin{equation}
    v=\sqrt{\frac{1}{1+ {r^2_{\rm ISCO}}/{\mathcal{L}^{2}_{\rm ISCO}}}} .
\end{equation}
Dependence of the ISCO position $r_{\rm isco}$ on the charge parameter ${\cal Q}$ and the change of the values of ${\cal L}_{\rm ISCO}$ and $v$ on the ISCO position are shown in Figure \ref{fig2}. ISCO is increasing for both positive and negative ${\cal Q}$. Similar results have been also obtained recently by \cite{2020GReGr..52...22H}, where the ISCO in a similar setting is properly discussed.

\begin{figure*}
\centering
\includegraphics[width=0.45\hsize]{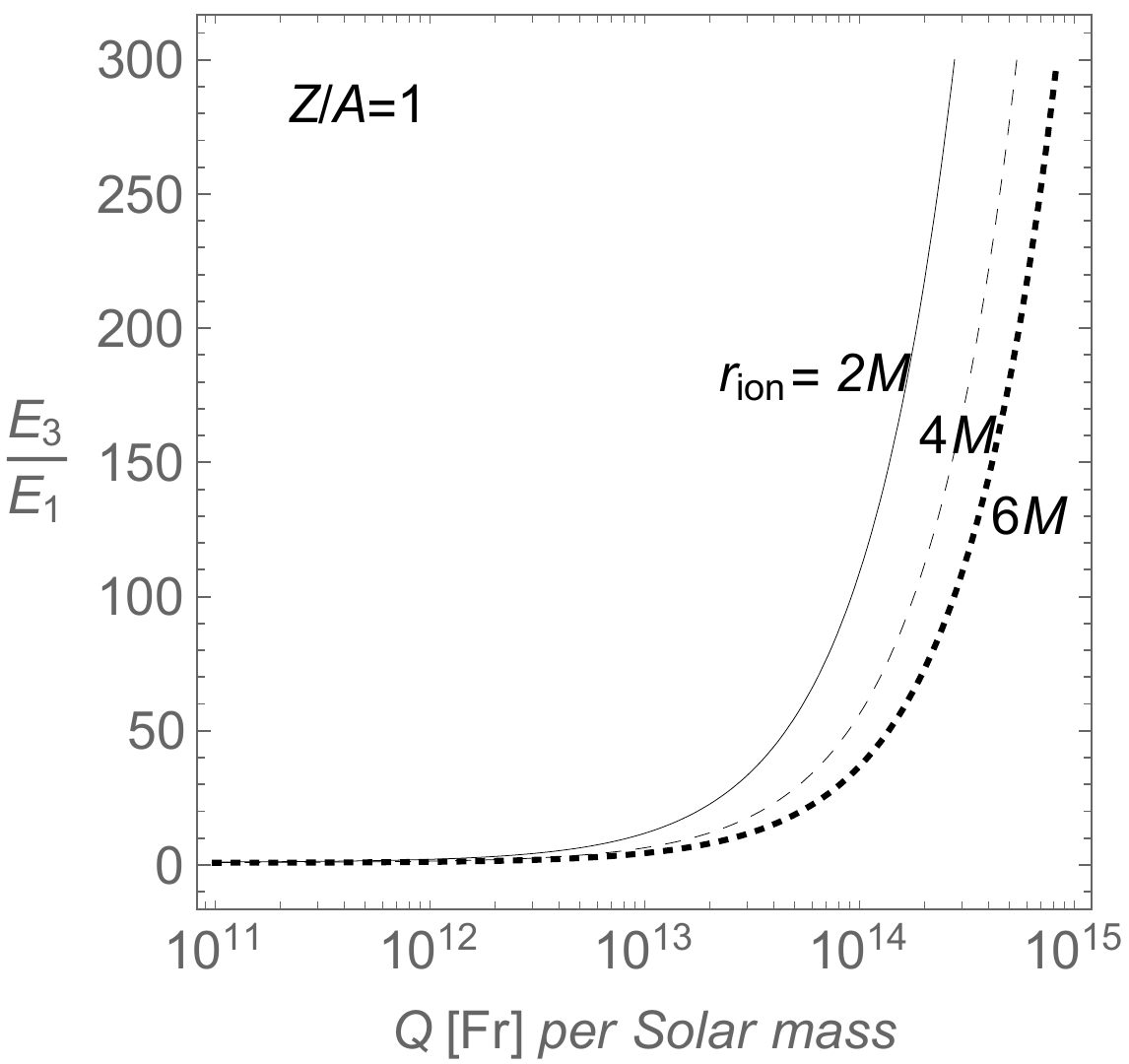}
\includegraphics[width=0.45\hsize]{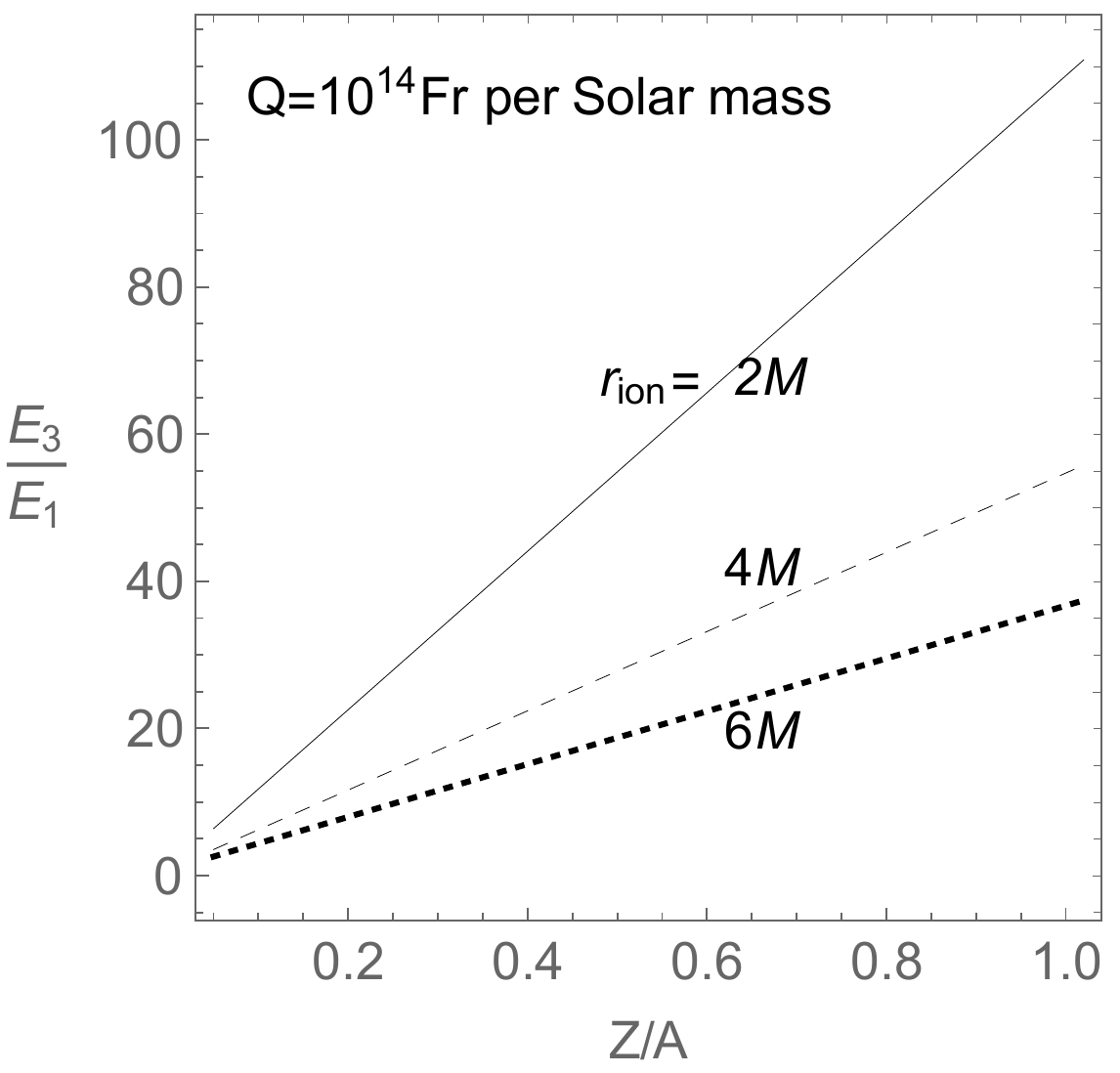}
\caption{Ratio of energies of ionized and neutral particles plotted against the black hole charge $Q$ (left) and the ionization rate $Z/A$ (right). The curves denote different positions of the ionization point: $r_{\rm ion}=2GM/c^2$ (solid), $r_{\rm ion}=4GM/c^2$ (dashed) and $r_{\rm ion}=6GM/c^2$ (dot-dashed). 
\label{fig3}}
\end{figure*}

\section{Energy of ionized particle}

\subsection{Conservation laws} 

Let us now consider the decay of a particle 1 into two fragments 2 and 3 close to the event horizon of a weakly charged Schwarzschild black hole at the equatorial plane. One can write the following conservation laws before and after decay
\begin{eqnarray}
E_{1} = E_{2} + E_{3} , \quad L_{1} = L_{2} + L_{3}, \quad q_{1} = q_{2}  + q_{3}, \\
m_1 \dot{r_{1}} = m_2 \dot{r_{2}} + m_3 \dot{r_{3}}, \quad m_1 \geqslant m_2 + m_3,
\end{eqnarray}
where dot indicates derivatives with respect to the particle's proper time $\tau$. 
Using above conservation laws, one can find the equation
\begin{equation}
m_1 u_{1} ^{\phi} = m_2 u^{\phi}_{2} + m_3 u^{\phi}_{3}. \label{conservation}
\end{equation}
Noticing that $u^{\phi} = \Omega u^{t} =  \Omega e / f(r)$, where $e_{i} = (E_{i} + q_{i} A_{t})/m_{i}$, with $i=1,2,3$ indicating the particle's number, the equation \eqref{conservation} will take the following from
\begin{equation}
\Omega_{1} m_{1} e_{1} = \Omega_{2} m_2 e_2 + \Omega_{3} m_3 e_3 .
\end{equation}
Solving the above equation with respect to the energy of one of the fragments, e.g. $E_3$ we find
\begin{equation}
E_{3} = \frac{\Omega_{1} - \Omega_{2}}{\Omega_{3} - \Omega_{2}} \left(E_1 + q_1 A_t\right) - q_3 A_t, 
\label{Energy3}
\end{equation}
where $\Omega_i = d \phi_i / d t$ is an angular velocity of $i$th particle, given by  (\ref{AngularVelocity}), with restricted values  (\ref{AngularVelocity:limits}).


\subsection{Maximum energy of ionized particle}

To maximize the energy of ionized particle we choose the particle 1 to be neutral, i.e. $q_1 = 0$. We are also free to choose the energy of the particle 1, which we set to its rest mass energy, i.e. $E_1 = m_1$ or ${\cal E} = 1$. In this case, the angular velocity (\ref{AngularVelocity}) for the particle 1 will take the following simple form
%
%
\begin{equation} \label{Omega1}
    \Omega_1 = \frac{1}{r^2} \sqrt{2(r-2)}.
\end{equation}
Without loss of generality we choose the ionized particle to be the particle 3. The energy of the ionized particle is maximal, when the term $(\Omega_{1} - \Omega_{2})/(\Omega_{3} - \Omega_{2})$ is maximized. This occurs when we set the angular momentum of fragments to their limiting values. Then we find
\begin{equation} \label{ratioOmega}
   \frac{\Omega_{1} - \Omega_{2}}{\Omega_{3} - \Omega_{2}} \Big |_{\rm max} = \frac{1}{\sqrt{2 ~r_{\rm ion}}}+\frac{1}{2},
\end{equation}
where $r_{\rm ion}$ is the ionization radius. 
We see that the ratio (\ref{ratioOmega}) decreases with increasing $r_{\rm ion}$ and maximal, when $r_{\rm ion}$ coincides with the event horizon. Thus, at $r_{\rm ion} = 2$, the ratio (\ref{ratioOmega}) is equal to unity. Finally, we write the expression for the energy of ionized particle in the form
\begin{equation}
    E_3 =  \left(\frac{1}{\sqrt{2 ~r_{\rm ion}}}+\frac{1}{2} \right) E_1 + \frac{q_3 Q}{ r_{\rm ion}} . 
\end{equation}

One can see that the energy of ionized particle is maximal when $q_3$ and $Q$ have the same sign, which is also expected result -- the charged particle is accelerated due to the Coulombic repulsion force acting between the black hole and particle. 
It is useful to define the ratio between the energies of ionized and neutral particles, which would represent the efficiency of the acceleration process. Writing the black hole mass and the speed of light explicitly and substituting $q_3 = Z e$ and $m_1 \approx A \,m_n$, where $Z$ and $A$ are the atomic and mass numbers, $e$ is an elementary charge and $m_n$ is the nucleon mass, we find
%
%
%
\begin{equation}
    \frac{E_3}{E_1} = \frac{1}{2} + 
    \sqrt{ \frac{G M}{2 \, c^2 \,r_{\rm ion}}}+ \frac{Z\, e\, Q}{A\, m_n \,c^2 \,r_{\rm ion}}. \label{EPP-effic}
\end{equation}
The ionized particle is accelerated only when the right hand side of the equation (\ref{EPP-effic}) is greater than unity. If the ionization point appears near the event horizon, $r_{\rm ion} \approx 2 G M /c^2$, then the condition $E_3 > E_1$ is satisfied for arbitrary positive values of the black hole charge, $Q>0$.  If the ionization point occurs at the ISCO radius, i.e. $r_{\rm ion} = 6 G M /c^2$, for the energy of ionized particle to be greater than the energy of infalling neutral particle, $E_3 > E_1$, the charge of the black hole has to satisfy the following condition  
\begin{equation}
 Q \gtrsim 5.8 \times 10^{11} \,\, \frac{A}{Z}  \,\,  \frac{M}{M_{\odot}} \,\,\, {\rm Fr} \, ,
\end{equation}
which is slightly greater than the lower limit of the estimated realistic limits of the black hole charge, given by (\ref{BHchargelimits}). 

In Figure \ref{fig3} (left) we plot the efficiency of the acceleration mechanism (ratio of the energies of ionized and infalling particles with $Z/A=1$) against the value of the black hole's charge per solar mass at various ionization points. The efficiency grows considerably with increasing the value of the black hole's charge and slightly decreases with increasing the distance between the black hole and ionization point. In Figure \ref{fig3} (right) we show the dependence of the energy ratio on the ionization rate $Z/A$ for a fixed value of the black hole charge. Let us estimate the maximal energy of ionized particle, which can be accelerated by the non-rotating weakly charged black hole. From (\ref{EPP-effic}) and using the uppermost realistic limit of the charge (\ref{BHchargelimits}) we get 
\begin{equation}
E_{\rm ion}^{\rm max} \approx 1.01 \times 10^6 \,\,\, Z \,\, \frac{Q}{10^{18} \, {\rm Fr}} \,\, \frac{M_\odot}{M} \,\, {\rm GeV},  \end{equation}
or, equivalently $\approx 1620$ erg. The ratio of energies of ionized and neutral particles, in this case, is equal to $E_{\rm ion}^{\rm max} / E_{n} \approx 10^{6}$. In sharp contrast to the magnetic Penrose process \citep{2020ApJ...895...14T,2019Univ....5..125T}, where the energy of ionized particle is increasing with increasing the black hole mass,  in the case of a non-rotating weakly charged black hole the energy of a charged particle is inversely proportional to the mass of the black hole. Therefore, the maximal energy is determined by the limiting value of the charge-to-mass ratio of the black hole $Q/M$ (see, the limits (\ref{BHchargelimits}) and the charge of the ionized particle $Z e$. This implies that the maximal energy of ionized particle accelerated by the weakly charged non-rotating black hole is similar for both stellar mass and supermassive black holes\footnote{Note that in the weakly magnetized rotating black hole case \cite{2019Univ....5..125T}, the energy of ionized particle grows proportionally to the black hole's mass and magnetic field strength $E_{\rm ion} \sim B M$.}.


\section{Possible application: high-energy cosmic rays}

\begin{figure}
\centering
\includegraphics[width=0.95\hsize]{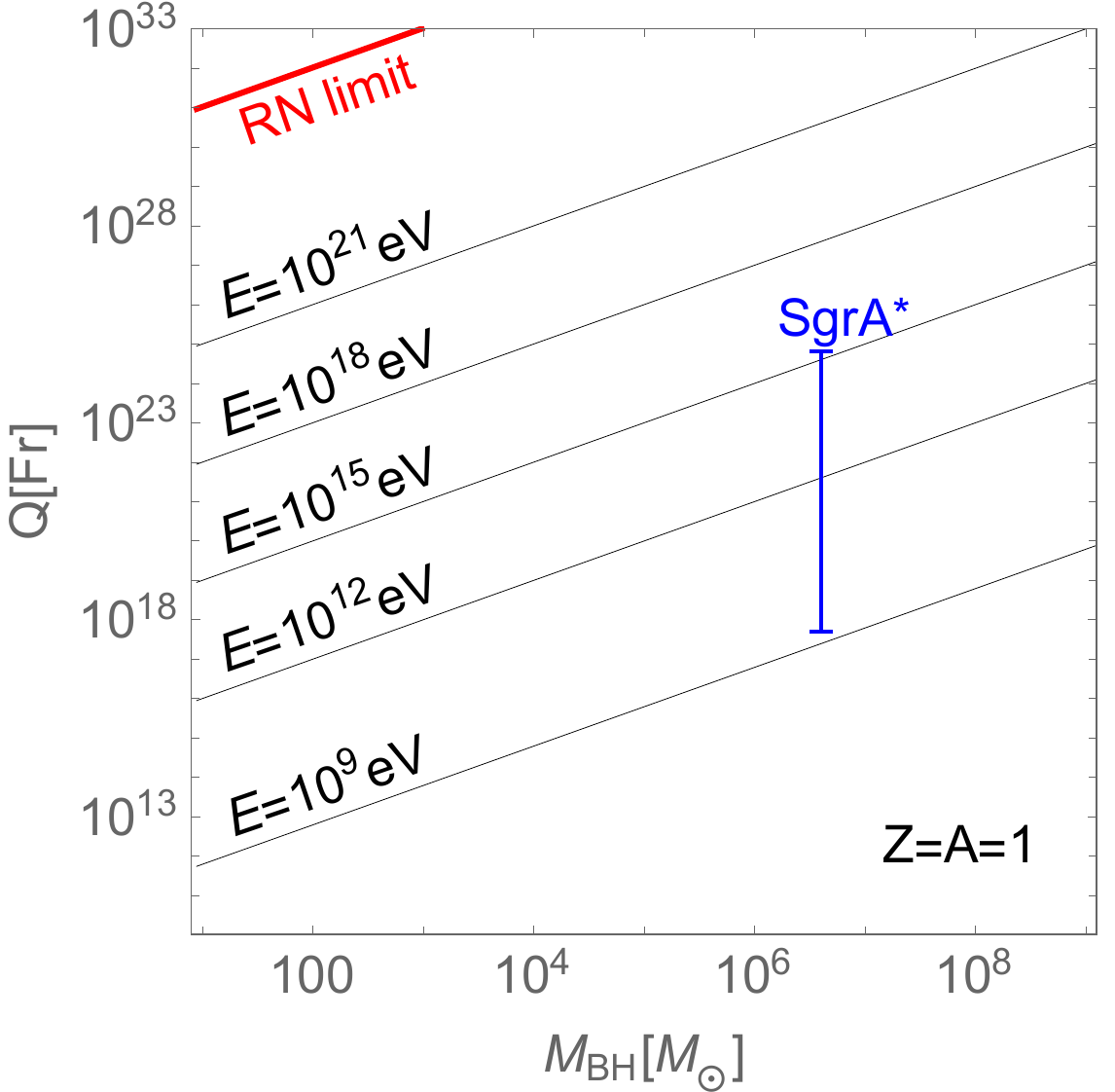}
\caption{Constraints on the black hole mass and its charge to  accelerate protons of various energies ($1-10^{12}$GeV), marked by black lines. The red line corresponds to the maximum theoretical limit of the Reissner-Nordstr\"om charge. The blue vertical line shows the acceleration capability of an example source, corresponding to a supermassive black hole Sgr A* located at the Galactic centre, whose charge constrains are taken from \cite{2018MNRAS.480.4408Z}. 
\label{fig4}}
\end{figure}

In \citep{2020ApJ...895...14T} it has been proposed that the ultra-efficient regime of the magnetic Penrose process, which requires the presence of an external magnetic field in the vicinity of rotating black hole can be relevant for the explanation of the origin of the ultra-high-energy cosmic rays (UHECRs). UHECRs are the phenomena composed from individual charged particles with detected energies exceeding $10^{20}$eV, whose production mechanism remains unknown. In this section we show that similar acceleration can be achieved in a more simplified setup of a weakly charged non-rotating black hole. In Figure~\ref{fig4} we demonstrate the constraints on the black hole mass and charge to serve as an accelerator of charged particles (protons) of certain energy. As it is expected, increasing the black hole charge for a given black hole mass one can reach the UHECR orders of energies. A central charge of the black hole, in this case, is still smaller than the maximal theoretical charge limit by many orders of magnitude. 

As an example, in Figure~\ref{fig4} we depict the acceleration capability of the Galactic centre supermassive black hole Sgr~A*, whose charge has been constrained in \cite{2018MNRAS.480.4408Z}. In particular, the constraint shows that Sgr~A* is capable to act as a PeVatron of charged particle with the energy of accelerated protons being of the order of $10^{15}$eV. It is interesting to note that this energy coincides with the knee of the cosmic ray spectrum,  above which the particle's flux is
considerably suppressed, suggesting the change in the cosmic ray source. The result is also similar to that obtained in the magnetic field and rotating black hole case \cite{2020ApJ...895...14T}, thus the presented mechanism can serve as another cosmic ray acceleration mechanism and alternative explanation of the cosmic ray knee when applied to the Galactic supermassive centre black hole Sgr~A*.

\section{Conclusions}

In this paper we proposed a simple mechanism of acceleration of particles to high-energies by the ionization of neutral particles in the vicinity of a non-rotating Schwarzschild black hole carrying small electric charge, whose gravitational 
effect on the spacetime metric is negligible. We 
started from the description of the motion of charged particle and showed that the effective potential in the case of a weakly charged black hole increases (or decreases) with increasing (or decreasing) the electric interaction parameter ${\cal Q}$. We found that the innermost stable circular orbit (ISCO) of the charged particle increases for both positive and negative values of the parameter ${\cal Q}$. The results are in accord with previous similar studies by \cite{2011PhRvD..83j4052P,2019Obs...139..231Z,2020GReGr..52...22H}. 

We have found that the energy of ionized particle can be much greater than initial energy of the neutral particle if both charges of the ionized particle and the black hole have the same sign. Thus, similar acceleration process occurring in the magnetized Kerr black hole spacetime and studied by \cite{2020ApJ...895...14T,2020Univ....6...26S} works also in the weakly charged non-rotating black hole case. For the realistic upper limit of the black hole charge given by (\ref{BHchargelimits}), which is at least 12 orders of magnitude smaller than the maximal theoretical Reissner-Nordstr\"om limit (\ref{RNcharge}), we have found that the charged particle with the charge $Z e$ can be accelerated to the Lorentz $\gamma$-factor exceeding $Z\times10^6$.

It is necessary to note that the energy of accelerated charged particles comes in expense of the electrostatic energy of the black hole given by its electric charge in contrast to many black hole energy extraction mechanisms, which use the rotational energy of the black hole (see, e.g. \cite{1969NCimR...1..252P,Bla-Zna:1977:MNRAS:,Wag-Dhu-Dad:1985:APJ:,2021PhRvD.103b4021K}). The electric Penrose process studied in the current paper is similar to the magnetic Penrose process, proposed in mid 1980's in \cite{Wag-Dhu-Dad:1985:APJ:,Par-etal:1986:APJ:,1985JApA....6...85B}, where it is the rotational energy of the black hole that is extracted. Connection of the magnetic Penrose process with another famous black hole energy extraction process -- the Blandford-Znajek mechanism \cite{Bla-Zna:1977:MNRAS:} was discussed in \cite{2018MNRAS.478L..89D}. It has been shown in \cite{2019Univ....5..125T} that the energy of accelerated ionized particle in the magnetic Penrose process is proportional to the product of the magnetic field strength and the black hole mass, i.e. $E_{\rm ion} \sim B \, M_{\rm BH}$. However, in the electric Penrose process studied in the current paper, this energy is proportional to the black hole's charge and inversely proportional to the black hole's mass, i.e. $E_{\rm ion} \sim Q / M_{\rm BH}$. Therefore, the maximal energy of ionized particle is restricted by the upper limit of the charge-to-mass ratio of a black hole. 


Further we also discussed possible astrophysical application of the presented results for the production and acceleration of the ultra-high-energy cosmic rays. In Figure~\ref{fig4} we presented the constraints on the maximum energy of the cosmic ray particle accelerated by a non-rotating black hole with given mass and charge. It has been shown that the ultra-high-energy acceleration does not require the presence of a strong black hole charge, in a sense that the Schwarzschild black hole spacetime is sufficient. We have applied the model to the Galactic centre supermassive black hole Sgr~A*, which is the best known black hole candidate, and found that the energy of accelerated protons can slightly exceed $10^{15}$eV, which coincides with the knee of the cosmic ray spectrum. The presented results, however, are quite general and can potentially operate also in neutron stars. It is especially interesting to look for a similar acceleration scenarios in the electrospheres of pulsars. Nevertheless, we leave this discussion for further studies. 

Despite the similarities between the acceleration processes described in the current paper and the magnetic Penrose process of particle acceleration by rotating black hole in a magnetic field \citep{2020ApJ...895...14T,2019Univ....5..125T}, one of the main differences between the two is that the motion of a charged particle in the former case is always regular, while in the later case the chaotic behaviour of escaping charged particle is usually observed \cite{2016EPJC...76...32S,2020Univ....6...26S}. In the presence of external magnetic field, it is expected that significantly larger number of charged particles escape along the the directions given by magnetic field lines. The character of the induced electric field around magnetized rotating Kerr black hole has no spherical symmetry, it is rather of quadrupole character. Meanwhile in the field of magnetized Schwarzschild black holes only redirections in chaotic motion are observed \cite{Kol-Stu-Tur:2015:CLAQG:}, but no acceleration is possible as no electric part of the field is induced. In a weakly charged non-rotating black hole case, the combined gravitational and electric field is spherically symmetric, therefore one would expect isotropic statistics of escaping charged particles with no preferred direction of the motion. In general, this can affect the statistics and interpretation of observed events, being of the distinguishing observational signature of the electric Penrose process.


\begin{acknowledgments}
The authors would like to acknowledge the Research
Centre for Theoretical Physics and Astrophysics and Institute of Physics of Silesian University in Opava for institutional support. Z. S. acknowledges the support of
the Grant No. 19-03950S of Czech Science Foundation
(GACR). 
\end{acknowledgments}



\providecommand{\noopsort}[1]{}\providecommand{\singleletter}[1]{#1}%
%



\end{document}